\documentclass{iopart}
\usepackage{iopams} 
\usepackage{graphics}
\usepackage{rotating}
\begin{document}

\title[]{Coherence features of the spin-aligned neutron-proton pair coupling scheme}

\author{C. Qi, J. Blomqvist, T. B\"ack, B. Cederwall, A. Johnson, R.J. Liotta, R. Wyss}

\address{Department of Physics, Royal Institute of Technology, SE-10691 Stockholm, Sweden}
\ead{chongq@kth.se}
\begin{abstract}
The seniority scheme has been shown to be extremely useful for the 
classification of nuclear states in semi-magic nuclei. The neutron-proton ($np$) correlation breaks the seniority symmetry in a major way.
As a result, the corresponding wave function is a mixture of many components with different seniority quantum numbers. In this contribution we show that the $np$ interaction may favor a new kind of coupling in $N=Z$ nuclei, i.e., the so-called isoscalar spin-aligned $np$ pair mode. Shell model calculations reveal that the ground and low-lying yrast states of the $N = Z$ nuclei $^{92}$Pd and $^{96}$Cd may mainly be built upon such spin-aligned $np$ pairs each carrying the maximum angular momentum $J = 9$ allowed by the shell $0g_{9/2}$ which is dominant in this nuclear region. 
\end{abstract}
The low-lying structures of many nuclei can be understood as the outcome of the competition between the pairing (or seniority) coupling scheme and the aligned coupling of individual particles in a non-spherical average potential. The seniority coupling dominates the low-lying states of semi-magic nuclei, 
where the driving force behind is the strong pairing interaction between like particles. The neutron-proton ($np$) correlation breaks the seniority symmetry in a major way.
Correspondingly, the wave function of open-shell system is a mixture of many components with different seniority quantum numbers. It is not clear yet how this kind of state can be classified in the $jj$-scheme. On the other hand, essential features of many open-shell nuclei can be described by assuming that the nucleons move in an average nuclear field by taking into account the fact that they have quadrupole moments that are  much larger than could be attributed to a single particle \cite{Bohr76}. From a spherical shell model point of view, the description of nuclear deformation involves the mixing of different configurations, in particular the coupling of quadrupole partners like $f_{7/2}$ and $p_{3/2}$ and $g_{9/2}$ and $d_{5/2}$.

Near the $N = Z$ line, nuclear collectivity
may be further enhanced by the interactions arising between neutrons and
protons occupying orbitals
with the same quantum numbers, which can form $np$ pairs with angular momenta $J=0$ to $2j$ and isospin quantum numbers $T=0$ (isoscalar) and $T=1$ (isovector). The isovector $np$ channel manifests itself in a fashion 
similar to like-nucleon correlations. In this contribution we would like to discuss the so-called spin-aligned $np$ pair coupling \cite{ced10,qi11} (See also \cite{Zer11}). In some spherical $N=Z$ nuclei like $^{92}$Pd (with four $np$ hole pairs inside $^{100}$Sn) and $^{96}$Cd (with two $np$ hole pairs) it may overcome the normal pairing as the dominant coupling scheme of the ground state structure \cite{ced10}. The aligned $np$ pairs can generate striking regular evolution patterns in the energy spectra and transition probabilities along the yrast states of $N=Z$ nuclei \cite{ced10,qi11}. These may be deemed as a new kind of collective mode with isoscalar character that is unique in the atomic nucleus.
Detailed calculations on the nucleus $^{92}$Pd can be found in \cite{ced10,qi11}. Here we will only concentrate on the simplest system of two neutrons and two protons in a single $j$ shell.

\begin{figure}
\begin{center}
\includegraphics[scale=0.3]{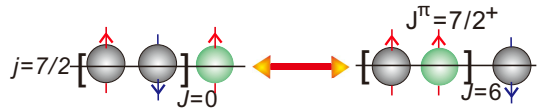}\\
\caption{A schematic picture for the coupling of three nucleons in a single $j=7/2$ shell.  These two schemes are the same for systems with three like nucleons. \label{seniority}
}
\end{center}
\end{figure}

Firstly we would like to remind briefly the basic idea behind the seniority scheme. As an illustration, in Figure \ref{seniority} we give a schematic plot for the coupling of three nucleons in a single $j=7/2$ shell. For a system with three like nucleons coupled to $J=j=7/2$, the state can be described as the product of a $J=0$ pair (with $v=0$) and an uncoupled nucleon. Thus it can be labeled as $v=1$. States with such configurations are usually lower in energy than others with high seniority numbers. The driving force behind this dominance of seniority coupling is the strong pairing interaction between like particles. But it has to be mentioned that
the lowest-seniority pair (with $v=0$) has nothing 
special from a coupling point of view since the nuclear state can then
be constructed in a variety of equivalent ways through other pairs. In 
particular, the aligned like-nucleon pair coupling was proposed in Ref. 
\cite{chen92} (see Figure \ref{seniority}). 
For the coupling of three like nucleons, these two schemes are equivalent since the $J=7/2$ state is uniquely defined. These two representation become non-equivalent only in systems with $j\geq9/2$. It is well known that seniority remains a good quantum number for systems with identical fermions in a single-$j$ shell when $j\leq7/2$. For shells with $j\geq 9/2$,  the rotationally-invariant interaction has to satisfy a certain number of linear constraints to conserve seniority. The number turns out to be [(2j-3)/6] ($[n]$ denotes the largest integer not exceeding $n$)~\cite{Talmi93,qi11a,qi10a}, which is related to the number of states with total spin $J=j$ for three identical nucleons in a single-$j$ shell. This indicates that the interactions that mix seniority are only a small fraction of the total two-body interactions~\cite{Talmi93}.

Figure \ref{seniority} also tells the  different faces of a quantum state. For a given state it can show to be a mixture of many different components in one representation while it may virtually be represented by a single coupling in another one, which should be deemed as a more proper representation. For the coupling of like nucleons, as mentioned before, this scheme was shown to be the seniority coupling. In the following we will show that it may correspond to a novel $np$ pair coupling when both protons and neutrons are involved.

For a simple system with $2n$ and $2p$ within a single $j$ 
shell, the wave function 
can be given as the tensor product of proton and neutron pair
components, i.e.,
\begin{equation}
|\Psi_I\rangle=\sum_{J_pJ_n}X_{J_pJ_n}|[\pi^2(J_p)\nu^2(J_n)]_I\rangle,
\end{equation}
where $J_p$ and $J_n$ denote the angular momenta of the proton and neutron pairs, respectively. They can only take even values with the maximum value $J=2j-1$. The amplitude $X$ is influenced by the $np$ interaction.  
As an example, in the hole-hole channel, the ground state wave function of $^{96}$Cd is calculated to be \cite{Qi10},
\begin{eqnarray}
|\Psi_0({\rm gs})\rangle&=& 0.76|[\pi^2(0)\nu^2(0)]_I\rangle + 0.57|[\pi^2(2)\nu^2(2)]_I\rangle \\
\nonumber &+&0.24 |[\pi^2(4)\nu^2(4)]_I\rangle+0.13|[\pi^2(6)\nu^2(6)]_I\rangle\\
\nonumber &+&0.14|[\pi^2(8)\nu^2(8)]_I\rangle.
\end{eqnarray}
It is seen that the normal monopole pair coupling scheme occupies only less than 60\% of the wave function.

This shell
model calculation was done in the standard fashion of using as representation
the tensorial product of neutron times proton degrees of freedom, thus easily
takes into account the Pauli principle. Thus we have $\langle[\pi^2(J'_p)\nu^2(J'_n)]_I|[\pi^2(J_p)\nu^2(J_n)]_I\rangle=\delta_{J'_pJ'_n}\delta_{J_pJ_n}$. The drawback with this representation
is that it is not straightforward to realize that the states are mainly determined by np pair degrees of freedom. This can only be done by projecting
the shell model wave function into the particular $np$ component one wishes
by using two-particle coefficients of fractional parentage.
One may re-express the wave function in an equivalent representation in terms of $np$ pairs. This can be done analytically with the help of the overlap matrix as
\begin{eqnarray}\label{over}
\nonumber\langle [\nu\pi(J_1)\nu\pi(J_2)]_I |[\pi^2(J_p)\nu^2(J_n)]_I\rangle \\
= \frac{-2}{\sqrt{N_{J_1J_2}}}\hat{J_1}\hat{J_2}
\hat{J_p}\hat{J_n}\left\{
\begin{array}{ccc}
j&j&J_p\\
j&j&J_n\\
J_1&J_2&I
\end{array}
\right\},
\end{eqnarray}
where $N$ denotes the normalization factor and $\hat{J}=\sqrt{2J+1}$. The overlap matrix automatically 
takes into account the Pauli principle.

\begin{figure}[htdp]
\begin{center}
\vspace{-0.5cm}
\includegraphics[scale=0.6,angle=270]{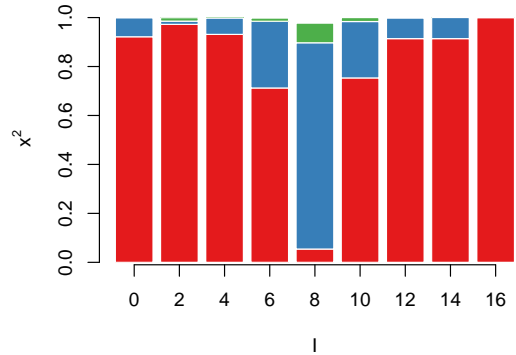}\\
\vspace{-0.5cm}
\caption{(Color online)
Coefficients $x^2$ corresponding to the $|((\nu\pi)_9)^2;0\rangle$ component in the wave functions of the first three $T=0$ states of $^{96}$Cd for each spin. The first, second and third states are denoted in red, blue and green colors, respectively.\label{str0}
}
\end{center}
\end{figure}

\begin{figure}[htdp]
\begin{center}
\includegraphics[scale=0.6]{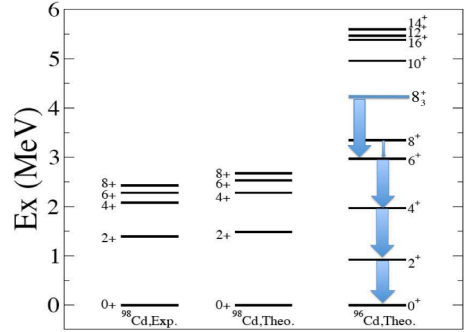}\\
\caption{Experimental \cite{Nudat} and calculated yrast level spectra of Cd isotopes. The width of the arrows indicates the E2 decay strength.\label{level}
}
\end{center}
\end{figure}

A striking feature thus found is that the calculated low-lying yrast states 
of $^{96}$Cd are mostly dominated by the coupling of spin-aligned $np$ pairs, as can be seen from Figure \ref{str0}. This may be compared with the stretch scheme proposed in \cite{dan67}.
An even more striking feature is that the low-lying yrast states are
calculated to be approximately equally spaced and their spin-aligned $np$ structure is the
same for all of them (up to spin $I=6$, see Figure \ref{level}).
Here the lower spin 
states (and also the highest spin states) are completely dominated by the aligned $np$ paired scheme  
$|((\nu\pi)_9\otimes(\nu\pi)_9)_I\rangle$. The situation becomes more complicated in states with spin around $I=8$ due to the competing effect from the isovector aligned pair $(0g_{9/2})^2_8$ (which is related to the kink in the calculated number of pairs along the yrast line \cite{Qi10}). In particular, the calculated $8^+_1$ state favors the non-collective seniority-like $((0g_{9/2})^2_0\otimes(0g_{9/2})^2_8)_{I=8}$ configuration rather than the aligned $np$ pair coupling. As a result, the $E2$ transition from this state to the $6^+_1$ state is hindered. Our calculations show that it is the third $8^{+}$ state (the second $T=0$, $8^+$ state) that corresponds to the spin-aligned $np$ pair coupling. 
 Moreover, the quadrupole transitions between these states show a
strong selectivity, since the decay to other structures beyond the np pair coupling scheme
is unfavored.
The calculated spectrum of $^{96}$Cd follows, as a function of 
the controlling parameter $\delta$, a pattern similar to that of 
$^{92}$Pd \cite{qi11}. The quantity $\delta$ leads to a fractional change of $V_9$, that is by taking as interaction matrix element
the value $V_9(\delta)= V_9(0)(1+\delta)$ where $V_9(0)$ denotes the one given in the original Hamiltonian \cite{qi11}. By enhancing the $(0g_{9/2})^2_9$, $T=0$, $J=9$ matrix element, a transition from the seniority coupling to the aligned $np$ pair coupling is seen in the $8^+_1$ state and  the equally-spaced pattern is extended to higher spins.

As one would expect, near the closed shell nucleus $^{100}$Sn, i.e., in
$^{96}$Pd and $^{98}$Cd, 
the positions of the energy levels correspond
to a $(g_{9/2})_\lambda ^2$ seniority
spectrum, as can be seen from Figure \ref{level}. 

It should be emphasized that only states with even angular momenta can be
generated from the spin-aligned $np$ coupling for systems with two pairs. The maximum spin
one can get is $2(2j-1)$. For each spin only one state can be formed by the coupling of two aligned pairs. 
The multiple pair $((\nu\pi)_9)^2_I$ can be written as a 
coherent superposition
of all isovector neutron and proton pairs occupying 
the $g_{9/2}$ shell in the 
form $((\nu^2_J)^{2} \otimes (\pi_J^2)^{2})_I$. For $I=0$ and $j=9/2$ we have
\begin{eqnarray}
|[\nu\pi(9)\nu\pi(9)]_0\rangle&=& 0.62|[\pi^2(0)\nu^2(0)]_I\rangle + 0.75|[\pi^2(2)\nu^2(2)]_I\rangle \\
\nonumber &+&0.23 |[\pi^2(4)\nu^2(4)]_I\rangle+0.02|[\pi^2(6)\nu^2(6)]_I\rangle\\
\nonumber &+&0.00|[\pi^2(8)\nu^2(8)]_I\rangle.
\end{eqnarray}

\begin{figure}[htdp]
\begin{center}
\includegraphics[scale=0.25]{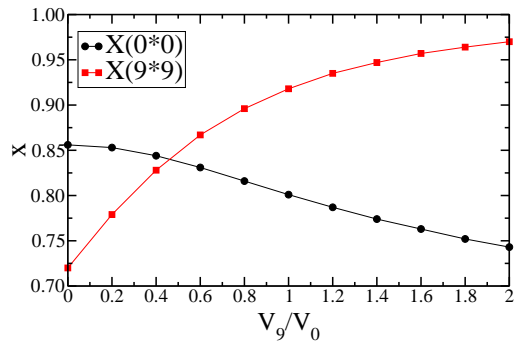}\\
\caption{Wave function amplitudes for the components $|(\nu\pi)_9\otimes(\nu\pi)_9\rangle$ and $|\nu^2_0\otimes\pi^2_0\rangle$ in the ground state wave function of $^{96}$Cd  as a function of the ratio between the strengths of $V_9$ and $V_0$. The calculation is done in the $0g_{9/2}$ shell with all other matrix elements being set to zero.\label{pair}
}
\end{center}
\end{figure}

To explore the role played by the $J=9$ interaction in inducing the $np$ pair coupling. In Figure \ref{pair} we calculated the wave function amplitudes for different components in the ground state wave function of $^{96}$Cd  with a schematic Hamiltonian containing only two terms $V_9$ and $V_0$. It is thus found that the aligned $np$ pairing coupling overcomes the normal monopole pair as the dominant component with $V_9/V_0\sim0.5$.

A spin trap with $I^{\pi}=16^+$ has been predicted in the four-hole system of $^{96}$Cd below the doubly magic $^{100}$Sn~\cite{Ogawa}.
This prediction is supported by our $0g_{9/2}$-shell calculations \cite{Qi10} as well as large-space calculations in the $fpg$-shell. In $j=9/2$ shell, the energy of this state  is,
\begin{equation}
E_{16}(^{96}{\rm Cd})=\frac{8}{17}V_7+3V_8+\frac{43}{17}V_9.
\end{equation}
This state can be constructed equivalently as the product of two aligned $np$ pair as $|(\nu\pi)_9\otimes(\nu\pi)_9\rangle$ and $|\nu^2_8\otimes\pi^2_8\rangle$.

In future works we plan to generalize the $np$ pair coupling to systems with many shells. Also we intend to understand the effects of such coupling on observables like the nuclear binding energy and radioactive decay properties.

This work has been supported by the Swedish Research Council (VR) under grant 
Nos. 623-2009-7340, 621-2010-3694 and 621-2010-4723.

\end{document}